\documentstyle[aps,pre,epsf]{revtex}

\begin{document}
\twocolumn[{\hsize\textwidth\columnwidth\hsize\csname
@twocolumnfalse\endcsname
\title{
\draft Effect of the Output of the System in Signal Detection
}
\author{
J. M. G. Vilar and J. M. Rub\'{\i}
}
\address{
Departament de F\'{\i}sica Fonamental, Facultat de
F\'{\i}sica, Universitat de Barcelona,\\
Diagonal 647, E-08028 Barcelona, Spain 
\date{\today}
}
\maketitle
\widetext
\begin{abstract}
\leftskip 54.8pt
\rightskip 54.8pt

We analyze the consequences that the choice of the output of the system
has in the efficiency of signal detection.
It is shown that the signal and the signal-to-noise ratio (SNR),
used to characterize the phenomenon of stochastic resonance,
strongly depend on the form of the output.
In particular, the SNR may be enhanced for an adequate output.

\end{abstract}
}]
\narrowtext

The phenomenon of stochastic resonance (SR)
\cite{Benzi,Maki1,Maki2,libi,JSP,Moss,Wies,Wiese,thre,phi4,mio1}
has emerged in the last years
as one of the most exciting in the field of nonlinear stochastic systems.
Its importance as a mechanism for signal detection has given
rise to a great number of applications in different fields,
as for example electronic devices \cite{tri},  lasers\cite{Maki1},
neurons \cite{neu1,neu2}  and  magnetic
particles \cite{Agus,grigo}.

The most common characterization of SR consists of the appearance of a
maximum in the output signal-to-noise ratio (SNR) at non-zero
noise level, although different definitions have been used in the literature.
The definition through the SNR accounts for practical applications,
because the SNR is the quantity that gives the
amount of information that
can be transfered through a medium as well as measuring
the quality of a signal. Additionally, the SNR
quantifies the possibility to  detect a signal embedded in a noisy
environment.
Another definition of SR, apparently similar to the one of the SNR, has
been proposed in terms of a maximum in the output signal.
Although both the SNR and the output signal
have been analyzed in terms of the parameters
of the system, e.g. the frequency or the amplitude of the input signal,
there is an important aspect which has not been considered in depth
up to now.
An adequate election of the output of the system may have implications in the
behavior of the quantities used to manifest the presence of SR.
This is precisely the problem we address in this paper.

It is interesting to realize that normally the output of the system
is the same as the dynamic variable $x(t)$ entering the stochastic
differential equation,
although sometimes the sign function of $x(t)$
has also been considered.
No matter the system, the output may in general be any function of $x(t)$
which is usually fixed through the characteristics of the problem.
However, in order to detect a signal embedded in a noisy environment
any function may be used. Thus, instead of Fourier transforming
$x(t)$ we can transform the function $v(x(t))$.

Let us discuss one of the most simplest cases, in which the dynamics
is described by an Ornstein-Uhlenbeck
process, where the input signal modulates the strength of the
potential in the following way:
\begin{equation} \label{model1}
{dx \over dt} = -h(t)x + \xi(t) \;\; .
\end{equation}
Here $h(t)=k(1+\alpha\sin(\omega_0 t))$,
with $k$ and $\alpha$  constants and  $\xi(t)$ is Gaussian white
noise with zero mean and second moment
$\left< \xi(t) \xi(t+ \tau) \right> = D \delta(\tau)$,
defining the noise level $D$.
The effect of this force may be
analyzed by the averaged power spectrum
\begin{equation} \label{ps}
P(\omega)={\omega_0 \over 2\pi}
\int_0^{2\pi/\omega_0}dt \int_{-\infty}^\infty
\left<v(t)v(t+\tau)\right>e^{-i\omega \tau}d\tau \;\; .
\end{equation}
To this end we will assume that it consists of a delta
function centered at the frequency $\omega_0$ plus a function $Q(\omega)$
which is smooth in the neighborhood of $\omega_0$ and is given by
\begin{equation} \label{ps2}
P(\omega)
%\equiv \vert \hat v(\omega) \vert^2
=Q(\omega)+S(\omega_0)\delta(\omega-\omega_0) \;\;.
\end{equation}
Let us now assume the explicit form for the output of the system,
$v(x)=\vert x \vert^\beta$, where $\beta$ is a constant.
Although this model does not exhibit SR, it is adequate to illustrate the
form in which signal and SNR vary as a function of the output.
Considerations about our model based upon dimensional analysis
enable us to rewrite the averaged power spectrum as
\begin{eqnarray} \label{mastereq}
P(\omega,D,k,\alpha,\omega_0,\beta)
&=& {1 \over k} \left({D \over k}\right)^\beta
q({\omega / \omega_0},{k / \omega_0},\alpha,\beta) \nonumber \\
&+& \left({D \over k}\right)^\beta
s({k / \omega_0},\alpha,\beta)
\delta\left(1-{\omega\over\omega_0}\right) \;\;,
\nonumber \\
& &
\end{eqnarray}
where $q({\omega / \omega_0},{k / \omega_0},\alpha)$
and $s({k / \omega_0},\alpha)$ are dimensionless functions.

\begin{figure}[th]
\centerline{
\epsfxsize=6cm 
\epsffile{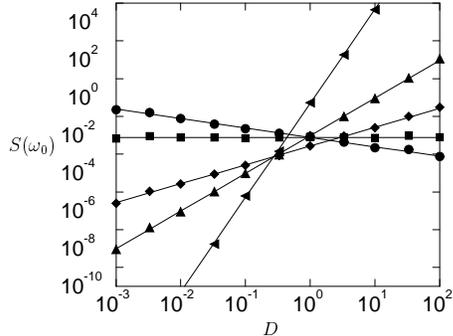}}
\caption[f]{\label{fig1}
Output signal corresponding to Eq. (\ref{model1})
($k=1$, $\alpha=0.5$, and $\omega_0/2\pi=0.1$) for different
exponents of the output ($\beta=-0.5,0,1,2,5$). The lines are
fit to a power law ($\beta_{fit}=-0.501,0.001,0.998,1.996,4.993$).
}
\end{figure}

In spite of the simplicity of this result,
a number of interesting consequences can be derived.
From Eq. (\ref{mastereq}) we can obtain the expression
for the output signal
\begin{equation}
{\rm S}=\left({D \over k}\right)^\beta s({k / \omega_0},\alpha,\beta) \;\; .
\end{equation}
Three qualitative different situations are present depending on the exponent
$\beta$. For $\beta>0$ the signal diverges when  the noise level $D$
goes to infinity, whereas for $\beta<0$ the signal
diverges when $D$ goes to zero.
Even more interesting is the case $\beta=0$, in which
the signal does not depend on the noise level.
The previous results have 
been verified numerically for some
particular values of $\beta$ (Fig. \ref{fig1}),
by integrating  the  corresponding  Langevin  equation following
a standard  second-order  Runge-Kutta  method for stochastic
differential  equations \cite{sde1,sde2}.

\begin{figure}[th]
\centerline{
\epsfxsize=6cm 
\epsffile{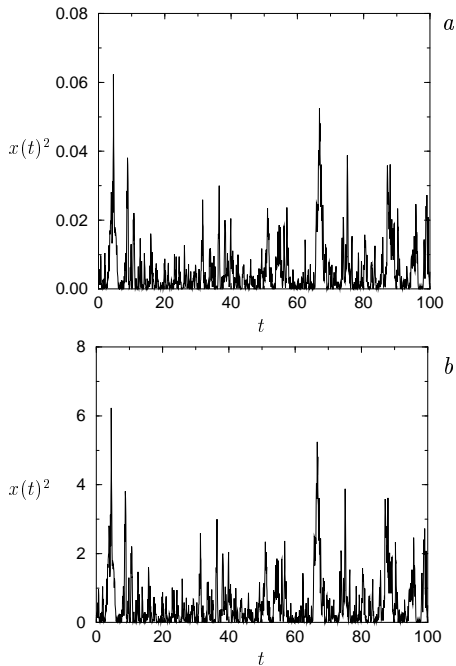}}
\caption[f]{\label{fig2}
Time evolution of $x(t)^2$ (same situation as in Fig. \ref{fig1})
for the noise levels (a) $D=0.01$ and (b) $D=1$.
}
\end{figure}

It is interesting to point out
that the signal increases
for low or high noise intensities, depending on the value of the 
exponent $\beta$.
From the previous considerations it becomes clear that 
the output signal itself does not always constitute a useful
quantity to elucidate the optimum noise level
to detect a signal.
In contrast, the SNR overcomes this ambiguity. Its expression
straightforwardly follows from Eq. (\ref{mastereq})
\begin{equation}
{\rm SNR}= k {s({k / \omega_0},\alpha,\beta)\over
q({\omega / \omega_0},{k / \omega_0},\alpha,\beta)} \;\; ,
\end{equation}
This result does not depend on the noise level thus 
indicating that the system is insensitive to the noise.
No matter the noise intensity, the SNR has always the same value
despite the fact that
signal is a monotonic increasing or decreasing function of
the noise.
For further illustration of these features
we have depicted in Fig. \ref{fig2}.
the temporal evolution of the output of the system
when $v(x)=x^2$, for two values of
the noise level. In both cases we have used the same
realization of the noise.
In the figure, we can see how the noise only affects the system
by changing its characteristic scales.

\begin{figure}[th]
\centerline{
\epsfxsize=6cm 
\epsffile{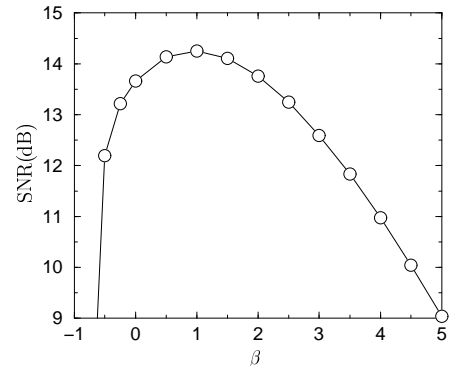}}
\caption[f]{\label{fig3}
SNR as a function of the exponent of the output
(same situation as in Fig. \ref{fig1}).
}
\end{figure}

The former results refer to the behavior of the SNR
as a function of $D$.
For practical applications
it is also
interesting the knowledge
of the SNR as a function of $\beta$
based upon the possible increasing of the SNR when varying $\beta$.
We have found that the SNR has a maximum at $\beta=1$ (see Fig. \ref{fig3}).
Consequently, for the output
class of functions $v(x)=\vert x \vert^\beta$ the input signal will be
better detected when $\beta=1$.

\begin{figure}[th]
\centerline{
\epsfxsize=6cm 
\epsffile{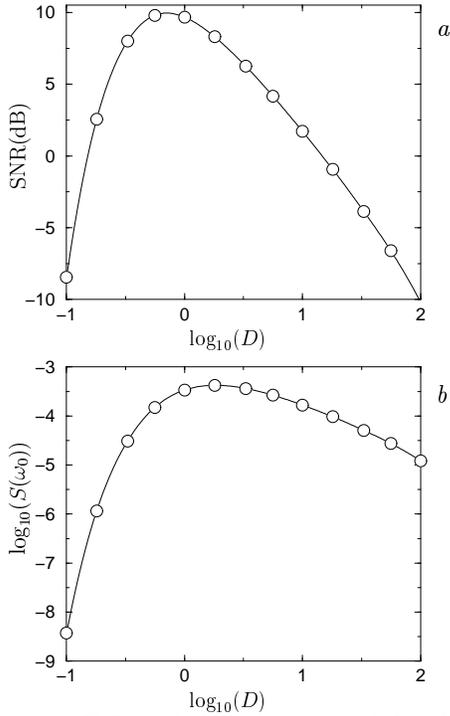}}
\caption[f]{\label{fig4}
(a) SNR and (b) output signal  for the output equal to the
step function with threshold $\theta=1$ 
(same situation as in Fig. \ref{fig1}).
}
\end{figure}

All the functions we are considering as outputs
are scale invariant, and dimensional analysis
can be readily performed. However,
when this requirement about $v(x)$ does not holds,
the previous results do not apply.
This could be the case of 
the Heaviside step function
$v(x)=\Theta(x-\theta)$, where $\theta$ represents a threshold.
In fact, this situation is quite similar to standard threshold
devices \cite{thre} considered previously.
In this case, both the SNR and the output signal exhibit a maximum at non-zero
noise level (see Fig. \ref{fig4}). Although
the evolution equation of the variable $x(t)$
is linear, SR appears due to the fact that the output
is a nonlinear function.

Having discussed the  role played by  the output in a simple monostable
system,
let us now analyze the case of the
bistable quartic potential, which has been frequently
proposed in order to describe the phenomenon of SR.
The dynamics of the system is then given by
the following equation:
\begin{equation} \label{lodesiempre}
{dx \over dt}=ax-bx^3 + A\sin(\omega_0t)+\xi(t)
\end{equation}
where a, b and $A$ are constants and $\xi(t)$
is the same noise as the one introduced through
Eq. (\ref{model1}).

To study this system, one usually takes as output the variable $x(t)$
and sometimes the sign function ${\rm sgn}(x(t))$.
In the limit when the amplitude of the input signal
goes to zero, these two forms of the output give the same
results (see ref. \cite{Maki2} for more details). However,
when the input signal has a finite amplitude, the SNR
for $x(t)$ diverges, whereas for ${\rm sgn}(x(t))$ goes to zero when the
noise level decreases. Despite the divergence of the SNR for $x(t)$, if the
amplitude of the input signal is not large enough, the SNR has a maximum at
non-zero $D$.

\begin{figure}[th]
\centerline{
\epsfxsize=6cm 
\epsffile{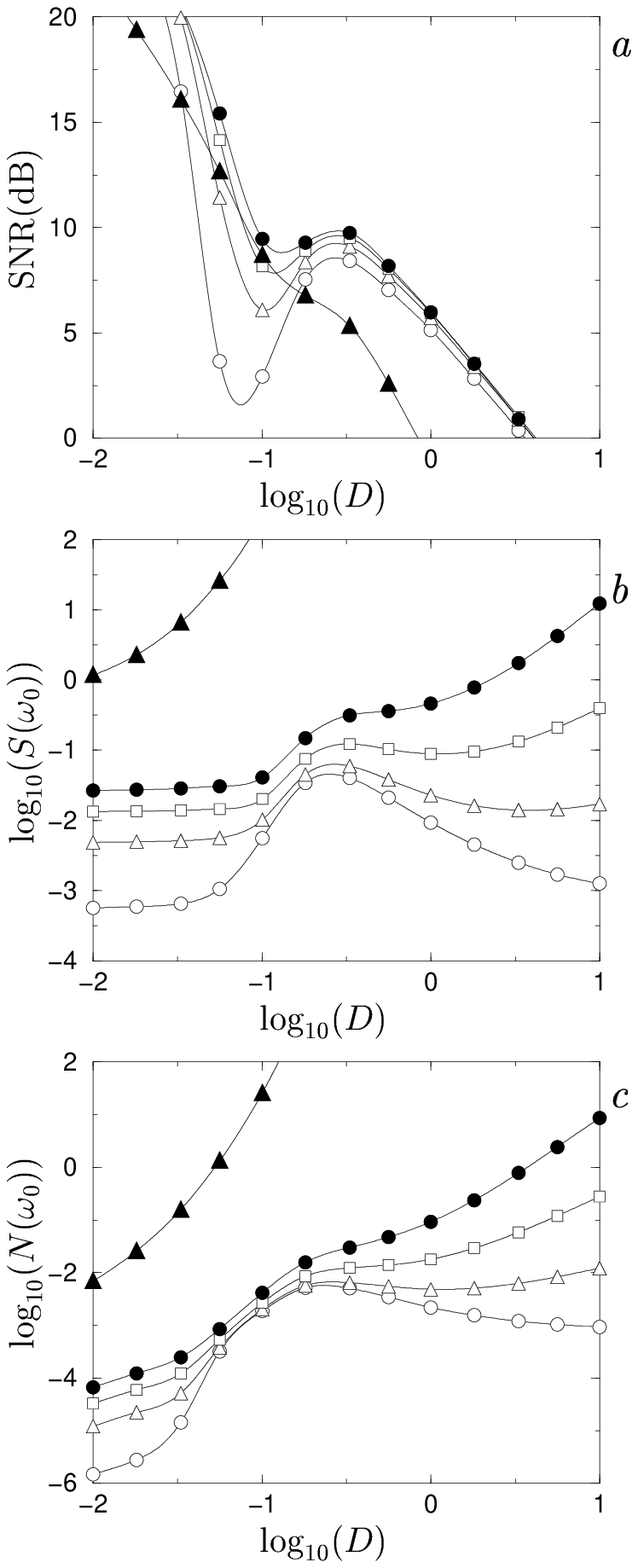}}
\caption[f]{\label{fig5}
(a) SNR, (b) output signal, and (c) output noise for the bistable
quartic potential ($a=1$, $b=1$, $\omega_0/2\pi=0.1$ and $A=0.13$)
for the outputs $\beta=1$ (empty circles), $\beta=3$ (empty triangles),
$\beta=5$ (empty squares), $\beta=7$ (filled circles),
and $\beta=25$ (filled triangles).
}
\end{figure}

As output,  we could take in general $x(t)^\beta$.
The choice of $\beta$ has important consequences as the SNR may
depend on this parameter. Thus to better detect a signal,
the noise level is not necessarily the only tunable
parameter.
In Fig. \ref{fig5}(a) we have plotted the SNR for different values of
$\beta$, observing its strong dependence on this parameter.
In particular, for $\log_{10}(D) \approx -1.25$,
upon varying $\beta$ from $1$ to $7$ the SNR 
increases in about 12 dB. Moreover, when increasing $\beta$ the 
maximum in the SNR becomes
less pronounced and disappears for a sufficiently large $\beta$, as
occurs for the case $\beta=25$ .
In regards to the signal,  variations of $\beta$ change
its behavior drastically.
This point is illustrated in Fig. \ref{fig5}(b) where we can see
that, when increasing the noise level,
for $\beta=1$ the signal goes to zero,
whereas for the remaining cases the signal always increases at
sufficiently high noise level.
In Fig. \ref{fig5}(c) we have also displayed the output noise.
From Fig. \ref{fig5}(a) it follows that a simple variation on the output
changes the qualitative form
of the SNR, in such a way that the maximum at non-zero noise level
may disappear.
Thus, the SNR is a monotonic decreasing function of $D$ and
apparently the input signal can always be better detected
by decreasing the noise level.
However, when
the SNR is a decreasing
function of $D$, 
there exists a region around the maximum,
corresponding to the curves $\beta=1,3,5,7$
in which the SNR for $\beta=1,3,5,7$ is greater than that for $\beta=25$.
We then conclude that when increasing the noise level, the signal can be better
detected if one simultaneously changes the value of $\beta$. 

\begin{figure}[th]
\centerline{
\epsfxsize=6cm 
\epsffile{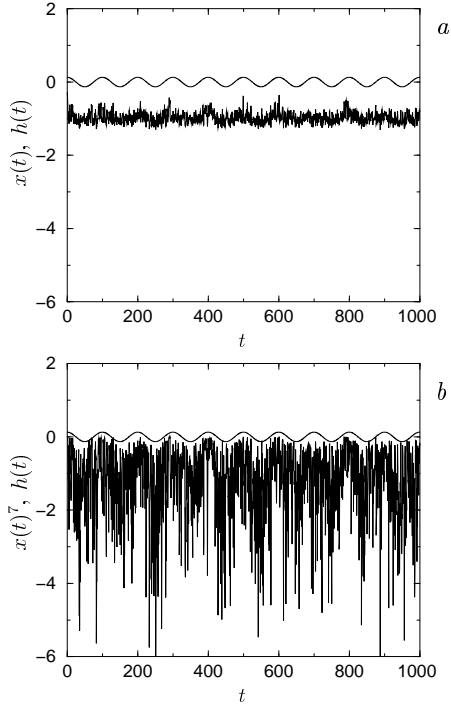}}
\caption[f]{\label{fig6}
Time evolution of (a) $x(t)$ and (b) $x(t)^7$
(same situation as in Fig. \ref{fig5}) for the noise
level $D=0.056$. The sinusoidal line in both figures
indicates the value of $h(t)$.
}
\end{figure}

To end our analysis, in Fig. \ref{fig6} we have displayed two temporal series
for two different values of $\beta$ at the noise level for which the effect
of the variation on $\beta$ is more pronounced.
We can see that intrawell oscillations, for $\beta=7$ are better
observed than for $\beta=1$.
This fact explains the increase of the SNR.

In summary,
we have shown that the quantities (signal and SNR)  used to characterize the 
phenomenon of SR strongly depend on the form of the output.
In this regard, the behavior of the SNR has revealed to be more robust
than the one corresponding to the signal.   
Our findings have important applied aspects since
an adequate choice of the output of the system
may be crucial in order to better detect a signal.

This work was
supported by DGICYT of the Spanish Government under Grant
No. PB96-0881. J.M.G.V. wishes to thank Generalitat
de Catalunya for financial support.

\end{document}